\documentclass[prd,showpacs,superscriptaddress,nofootinbib,twocolumn,twoside]{revtex4}
\usepackage[utf8]{inputenc}

\newcommand{\bea}{\begin{eqnarray}}
\newcommand{\eea}{\end{eqnarray}}
\newcommand{\bes}{\begin{subequations}}
\newcommand{\ees}{\end{subequations}}

\usepackage[paperwidth=210mm,paperheight=297mm,centering,hmargin=2cm,vmargin=2cm]{geometry}
\usepackage{mathrsfs,mathtools,bbold}
\usepackage{amsmath,amssymb}
\usepackage{amsfonts}
\usepackage{verbatim}
\usepackage{graphicx}
\usepackage{mathrsfs,mathtools}
\usepackage{ccaption}
\usepackage[dvipsnames]{xcolor}
\definecolor{SchoolColor}{rgb}{0.6471, 0.1098, 0.1882} % Crimson
\usepackage{stmaryrd}
\usepackage{textcomp}
\usepackage[colorlinks=true,linkcolor=black,citecolor=teal,urlcolor=MidnightBlue,filecolor=black]{hyperref}

\newcommand{\unity}{1\hspace{-0.243em}\text{l}}

\newcommand{\be}[1]{ \begin{equation}\label{#1} }
\newcommand{\ee}{\end{equation}}

\DeclareMathOperator{\extdm}{d}
\newcommand{\extd}{\extdm \!}

\usepackage{ulem} % For commenting commands
\usepackage[makeroom]{cancel}

\definecolor{color1}{rgb}{0.22,0.45,0.70}

\begin{document}

%---------------------------------------------------------------------------------------------------------------
\title{Enhanced Asymptotic Symmetry Algebra of 2+1 Dimensional Flat Space}
%---------------------------------------------------------------------------------------------------------------

\author{St{\'e}phane Detournay}
\email{sdetourn@ulb.ac.be}

\author{Max Riegler}
\email{Max.Riegler@ulb.ac.be}

\affiliation{Universit{\'e} libre de Bruxelles, Boulevard du Triomphe (Campus de la Plaine), 1050 Bruxelles, Belgium, Europe}

%\date{\today}

\begin{abstract}
In this paper we present a new set of asymptotic boundary conditions for Einstein gravity in 2+1 dimensions with vanishing cosmological constant that are a generalization of the Barnich-Comp{\`e}re boundary conditions \cite{Barnich:2006av}. These new boundary conditions lead to an asymptotic symmetry algebra that is generated by a $\mathfrak{bms}_3$ algebra and two affine $\hat{\mathfrak{u}}(1)$ current algebras. We then apply these boundary conditions to Topologically Massive Gravity (TMG) and determine how the presence of the gravitational Chern-Simons term affects the central extensions of the asymptotic symmetry algebra. We furthermore determine the thermal entropy of solutions obeying our new boundary conditions for both Einstein gravity and TMG.
\end{abstract}

\pacs{04.20.Ha,04.60.Rt,11.15.Yc,04.70.Dy}

\maketitle

\tableofcontents
\hypersetup{linkcolor=SchoolColor}

%---------------------------------------------------------------------------------------------------------------
\section{Introduction}
%---------------------------------------------------------------------------------------------------------------

Einstein gravity in 2+1 dimensions has many features that distinguish it from higher-dimensional theories of gravity. Maybe the most striking difference is the absence of local propagating degrees of freedom. Thus in Einstein gravity only global effects are of physical relevance. As such, it is of great interest to find boundary conditions that lead to interesting boundary dynamics.\\
In the case of Anti-de Sitter (AdS) spacetimes in 2+1 dimensions Brown and Henneaux showed in a seminal paper \cite{Brown:1986nw}, that by choosing suitable boundary conditions one can enlarge the $\mathfrak{sl}(2,\mathbb{R})\oplus\mathfrak{sl}(2,\mathbb{R})$ bulk isometries to two copies of a Virasoro algebra with central charges $c=\bar{c}=\frac{3\ell}{2G_N}$, where $\ell$ is the AdS radius and $G_N$ is Newton's constant. These boundary conditions also include BTZ black holes \cite{Banados:1992wn,Banados:1992gq} and thus cover a wide range of physical applications.\\
As such, these boundary conditions were modified and generalized in many ways e.g.~in the presence of scalar matter \cite{Henneaux:2002wm}, a gravitational Chern--Simons (CS) term \cite{Grumiller:2008es, Henneaux:2009pw, Skenderis:2009nt, Afshar:2011qw} or other higher derivative interactions \cite{Hohm:2010jc, Sinha:2010ai}.\\
In addition several alternatives to the Brown-Henneaux boundary conditions were discovered yielding asymptotic symmetry algebras that differ from the conformal algebra in two dimensions \cite{Compere:2013bya,Troessaert:2013fma,Avery:2013dja,Troessaert:2015gra,Donnay:2015abr,Afshar:2015wjm,Afshar:2016wfy,Perez:2016vqo,Setare:2016jba,Setare:2016vhy,Grumiller:2016pqb}.\\
Almost twenty years after Brown and Henneaux, Barnich and Comp{\`e}re \cite{Barnich:2006av} presented a consistent set of boundary conditions for asymptotically flat spacetimes at null infinity\footnote{For a CS version of these boundary conditions see e.g. \cite{Barnich:2013yka,Afshar:2013bla}.} that extended previous considerations of \cite{Ashtekar:1996cd}. Using these boundary conditions Barnich and Comp{\`e}re were able to show that the corresponding asymptotic symmetry algebra is given by the three dimensional Bondi-Metzner-Sachs algebra ($\mathfrak{bms}_3$) \cite{Bondi:1962px,Sachs:1962zza}. Similar to the Brown-Henneaux boundary conditions in the AdS$_3$ case the Barnich-Comp{\`e}re boundary conditions also include a class of cosmological solutions called "flat space cosmologies" (FSCs) \cite{Cornalba:2002fi,Cornalba:2003kd}. However, in contrast to the AdS$_3$ case there have been much fewer\footnote{To the best of our knowledge there are only the boundary conditions in \cite{Afshar:2015wjm} that support asymptotic Rindler spacetimes as well as the boundary conditions in \cite{Afshar:2016kjj} that feature two copies of a Heisenberg algebra as asymptotic symmetries.} alternative boundary conditions for asymptotically flat spacetimes in 2+1 dimensional Einstein gravity aside from the Barnich-Comp{\`e}re boundary conditions up until now.\\
In this paper we will improve that situation by providing a novel set of boundary conditions for asymptotically flat spacetimes in 2+1 dimensional Einstein gravity. The resulting asymptotic symmetry algebra will be given by a semidirect product of a $\mathfrak{bms}_3$ algebra with central charge $c_M=\frac{1}{4G_N}$ and two affine $\hat{\mathfrak{u}}(1)$ current algebras with identical levels $\kappa_P=-\frac{1}{4G_N}$. These boundary conditions can also be viewed as the flat space counterpart of the enhanced symmetry algebra for AdS$_3$ found in \cite{Troessaert:2013fma} \\
The organisation of the paper is as follows. In Section \ref{sec:CS} we present the new set of boundary conditions using the Chern-Simons formulation of gravity and perform a canonical analysis that determines the form of the corresponding asymptotic symmetry algebra. In section \ref{sec:TMG} we first translate our new boundary conditions into a second order formulation and then subsequently apply these boundary conditions to Topologically Massive Gravity (TMG) \cite{Deser:1981wh,Deser:1982vy}. In Section \ref{sec:ThermalEntropy} we determine the thermal entropy of the solutions that are allowed by our boundary conditions. We then show in Section \ref{eq:Troessaert} how one can relate our boundary conditions to the ones presented in \cite{Troessaert:2013fma} as a limiting case of vanishing cosmological constant.

%---------------------------------------------------------------------------------------------------------------
\section{Boundary Conditions and Asymptotic Symmetries}\label{sec:CS}
%---------------------------------------------------------------------------------------------------------------

In this section we present new boundary conditions for asymptotically flat spacetimes in 2+1 dimensional Einstein gravity and derive the corresponding asymptotic symmetry algebra using the Chern-Simons formulation of gravity.

%---------------------------------------------------------------------------------------------------------------
\subsection{Chern-Simons Action and Conventions}
%---------------------------------------------------------------------------------------------------------------

Our starting point is the Chern-Simons formulation of gravity in 2+1 dimensions \cite{Witten:1988hc}. In this formulation the gravitational dreibein $e$ and the associated (dualized) spin connection $\omega$ are recombined into a new gauge field $\mathcal{A}=\omega+e$. The Einstein-Hilbert action in the Palatini formulation is then equivalent (up to boundary terms) to a Chern-Simons action of the form
	\begin{equation}\label{eq:CSFS}
		S_{\rm CS}[\mathcal{A}] = \frac{k}{4\pi} \int_{\mathcal{M}} \langle\mathcal{A} \wedge \extd \mathcal{A} +\frac23 \mathcal{A} \wedge \mathcal{A} \wedge \mathcal{A}\rangle\,,
	\end{equation}
where $\langle\ldots\rangle$ is an appropriate invariant bilinear form on a given gauge algebra, $k$ is the Chern-Simons coupling and $\mathcal{M}$ a 2+1-dimensional manifold. Depending on the kind of geometry one wants to describe the gauge field $\mathcal{A}$ takes its values in certain Lie algebras. For gravity theories with vanishing cosmological constant the appropriate Lie algebra to use is $\mathfrak{isl}(2,\mathbb{R})$ for which we use the following basis:
	\begin{subequations}\label{eq:isl2RBasis}
	\begin{align}
		[L_n,L_m]&=(n-m)L_{n+m},\\
		[L_n,M_n]&=(n-m)M_{n+m},\\
		[M_n,M_m]&=0,
	\end{align}
	\end{subequations}
where $n,m=0,\pm1$ and the corresponding invariant bilinear form is given by $\langle L_nL_m\rangle=\langle M_nM_m\rangle=0$ as well as
	\begin{equation}\label{eq:ISLInvBilForm}
		\langle L_nM_m\rangle=-2\left(
			\begin{array}{c|ccc}
				  &M_1&M_0&M_{-1}\\
				\hline
				L_1&0&0&1\\
				L_0&0&-\frac{1}{2}&0\\
				L_{-1}&1&0&0
			\end{array}\right).
	\end{equation}
In order to make contact to the Einstein-Hilbert-Palatini action for vanishing cosmological constant the Chern-Simons coupling $k$ has to be related to Newton's constant $G_N$ in 2+1 dimensions by
	\begin{equation}
		k=\frac{1}{4G_N}.
	\end{equation}
This is the basic setup that we use to define our boundary conditions.

%---------------------------------------------------------------------------------------------------------------
\subsection{Boundary Conditions}
%---------------------------------------------------------------------------------------------------------------

Before imposing any kind of boundary conditions one has to further specify the topology of the manifold on which one is imposing said boundary conditions. For this work we are assuming that the topology of the manifold is given by a solid cylinder. In addition we choose coordinates such that there is a radial direction $0\leq r<\infty$ and the boundary of the cylinder is parametrized by a retarded time coordinate $-\infty<u<\infty$ as well as an angular coordinate $\varphi\sim\varphi+2\pi$.\\
One advantage of the Chern-Simons formulation of gravity is that one can use some of the gauge freedom to fix the radial dependence of gauge field $\mathcal{A}$ as
	\begin{equation}
		\mathcal{A}(r,u,\varphi)=b^{-1}(r)\left[a(u,\varphi)+\extd\,\right]b(r),	
	\end{equation}
with
	\begin{equation}
		a(u,\varphi)=a_\varphi(u,\varphi)\extd\varphi+a_u(u,\varphi)\extd u.
	\end{equation}	
A consequence of such a gauge choice is that the equations of motion $F=\extd\mathcal{A}+[\mathcal{A},\mathcal{A}]=0$ simplify drastically. Except $F_{u\varphi}$ all components of $F_{\mu\nu}$ are identically zero for such a choice of gauge.\\
It is important to note that different choices of the group element $b$ will yield different geometrical interpretations. In order to interpret our boundary conditions as cosmological solutions later on  we choose this group element as
	\begin{equation}
		b(r)=e^{\frac{r}{2}M_{-1}}.
	\end{equation}
After having completely specified our specific setup we are now ready to formulate boundary conditions. Thus we propose the following new boundary conditions for asymptotically flat spacetimes in 2+1 dimensions:
	\begin{subequations}\label{eq:NewFSBCs}
	\begin{align}
		a_\varphi&=e^\alpha L_1+\bar{\mathcal{M}}L_{-1}+\bar{\beta}M_1+\bar{\mathcal{N}}M_{-1},\\
		a_u&=e^\alpha M_1+\bar{\mathcal{M}}M_{-1},
	\end{align}
	\end{subequations}
where the functions $\bar{\mathcal{M}}$, $\bar{\mathcal{N}}$, $\alpha$ and $\bar{\beta}$ are arbitrary functions of $u$ and $\varphi$.\\
In principle one could perform the asymptotic analysis using the boundary conditions as written in \eqref{eq:NewFSBCs}. However, with a bit of hindsight it turns out that the following redefinitions will be beneficial to determine the asymptotic symmetries:
	\begin{equation}
		\bar{\mathcal{M}}=-\frac{\mathcal{M}}{4}e^{-\alpha},\,\,\bar{\beta}=\beta e^\alpha,\,\, \bar{\mathcal{N}}=\frac{e^{-\alpha}}{4}(\mathcal{M}\beta-2\mathcal{N}).
	\end{equation}
Looking at the equations of motion $F=0$ one can already see that this redefinition is very useful as it yields very simple constraints on the time evolution of the functions $\mathcal{M}$, $\mathcal{N}$, $\alpha$ and $\beta$ as
	\begin{equation}\label{eq:EOM}
		\partial_u\alpha=\partial_u\mathcal{M}=0,\,\partial_u\beta=\partial_\varphi\alpha,\, 2\partial_u\mathcal{N}=\partial_\varphi\mathcal{M}.
	\end{equation}
That means that on-shell these functions can be written as
    \begin{subequations}\label{eq:StateOnShell}
    \begin{align}
    \mathcal{M} &=  \mathcal{M}(\varphi),& \mathcal{N} &= \mathcal{L}(\varphi) +\frac{u}{2}\mathcal{M}' \\
    \alpha &= A(\varphi),&\beta &= B(\varphi) + u A'.
    \end{align}
    \end{subequations}
%---------------------------------------------------------------------------------------------------------------
\subsection{Variational Principle}
%---------------------------------------------------------------------------------------------------------------

A first consistency check of the boundary conditions \eqref{eq:NewFSBCs} is to see whether or not they lead to a well defined variational principle. This check can be performed by varying the Chern-Simons action \eqref{eq:CSFS} with respect to the gauge field $\mathcal{A}$ and plugging in the boundary conditions \eqref{eq:NewFSBCs}. If the resulting expression vanishes on-shell one has a well defined variational principle. If that is not the case one has to try and look for suitable boundary terms that can be added to the action to make the variation vanish on-shell.\\
Varying the Chern-Simons action \eqref{eq:CSFS} yields
	\begin{equation}\label{eq:CheckVarPrinciple}
		\delta S_{\rm CS}[\mathcal{A}]=\frac{k}{2\pi}\int_\mathcal{M}\langle\delta\mathcal{A}\wedge F\rangle+\frac{k}{4\pi}\int_{\partial\mathcal{M}}\langle\delta\mathcal{A}\wedge\mathcal{A}\rangle.
	\end{equation}
The first term vanishes for $F=0$ so only the second term could be potentially problematic and require an additional counterterm. However, one can explicitly check that for the boundary conditions \eqref{eq:NewFSBCs} this term vanishes. Thus the variation of the Chern-Simons action is zero on-shell and the proposed boundary conditions yield a consistent variational principle.

%---------------------------------------------------------------------------------------------------------------
\subsection{Boundary Condition Preserving Gauge Transformations}
%---------------------------------------------------------------------------------------------------------------

After having checked that the boundary conditions lead to a well defined variational principle one has to determine the gauge transformations $\delta_\epsilon\mathcal{A}=\extd\epsilon+[\mathcal{A},\epsilon]$ that preserve the boundary conditions \eqref{eq:NewFSBCs}.
These gauge transformations contain two different sets of transformations, referred to as proper and improper. The latter have an associated non-trivial canonical boundary charge, hence act non-trivially at the boundary and change the state of the theory. The asymptotic symmetries of a given set of boundary conditions are given by the set of gauge transformations that act non-trivially at the boundary modulo the proper gauge transformations.\\
In order to find the set of asymptotic symmetries we make the ansatz
	\begin{equation}\label{eq:GaugeParameter}
		\epsilon(\rho,u,\varphi)=b^{-1}\left[\sum\limits_{a=-1}^1\epsilon^a(u,\varphi)L_a+\sigma^a(u,\varphi)M_a\right]b.
	\end{equation}
In terms of this ansatz the gauge transformations (including proper and non-trivial ones) that preserve the boundary conditions \eqref{eq:NewFSBCs} are given by
	\begin{subequations}\label{eq:BCPGTs}
	\begin{align}
		\epsilon^1&=e^\alpha\epsilon_\mathcal{L},\quad \epsilon^0=\epsilon_\mathcal{J},\nonumber\\
		\epsilon^{-1}&=-\frac{e^{-\alpha}}{4}\left(2\epsilon_\mathcal{J}'+\mathcal{M}\epsilon_\mathcal{L}\right),\\
		\sigma^1&=e^\alpha\left(\beta\epsilon_\mathcal{L}+\sigma_\mathcal{M}\right),\quad\sigma^0=\sigma_\mathcal{P},\nonumber\\
		\sigma^{-1}&=\frac{e^{-\alpha}}{2}\left[\frac{\mathcal{M}}{2}\left(\beta\epsilon_\mathcal{L}-\sigma_\mathcal{M}\right)-\mathcal{N}\epsilon_\mathcal{L}+\beta\epsilon_\mathcal{J}'-\sigma_\mathcal{P}'\right],
	\end{align}
	\end{subequations}
where the functions $\epsilon_\mathcal{L}$, $\epsilon_\mathcal{J}$, $\sigma_\mathcal{M}$ and $\sigma_\mathcal{P}$ depend on $u$ and $\varphi$ and a prime denotes differentiation with respect to $\varphi$. In addition these functions have to satisfy
	\begin{equation}\label{eq:GaugeParametersTimeEvolution}
		\partial_u\epsilon_\mathcal{L}=\partial_u\epsilon_\mathcal{J}=0,\,\partial_u\sigma_\mathcal{M}=\partial_\varphi\epsilon_\mathcal{L},\,\partial_u\sigma_\mathcal{P}=\partial_\varphi\epsilon_\mathcal{J}.
	\end{equation}
That means that these gauge parameters can also be written as
    \begin{subequations}
    \begin{align}
    \epsilon_\mathcal{L} &=  \epsilon_\mathcal{L}(\varphi),& \sigma_\mathcal{M} &= \epsilon_\mathcal{M}(\varphi) +u\epsilon_\mathcal{L}' \\
    \epsilon_\mathcal{J} &= \epsilon_\mathcal{J}(\varphi),&\sigma_\mathcal{P} &= \epsilon_\mathcal{P}(\varphi) + u \epsilon_\mathcal{J}'.
    \end{align}
    \end{subequations}
In order to distinguish proper gauge transformations from non-trivial ones we now determine the canonical boundary charges. Again with a bit of hindsight we first redefine the functions $\mathcal{L}$, $\mathcal{M}$, $A$ and $B$ that are shown in \eqref{eq:StateOnShell} as
	\begin{subequations}\label{eq:RedefState}
	\begin{align}
		\tilde{\mathcal{M}}&=\frac{k}{4\pi}\mathcal{M},&\tilde{\mathcal{L}}&=\frac{k}{2\pi}\mathcal{L},\\\mathcal{P}&=-\frac{k}{2\pi}A',&\mathcal{J}&=-\frac{k}{2\pi}B'.
	\end{align}
	\end{subequations}
The redefined fields $\tilde{\mathcal{M}}$, $\tilde{\mathcal{L}}$, $\mathcal{J}$ and $\mathcal{P}$ then transform under the gauge transformations \eqref{eq:BCPGTs} as
	\begin{subequations}\label{eq:GaugeTrafos}
	\begin{align}
		\delta_\epsilon\tilde{\mathcal{M}}&=\epsilon_\mathcal{L}\tilde{\mathcal{M}}'+2\tilde{\mathcal{M}}\epsilon_\mathcal{L}'+\mathcal{P}\epsilon_\mathcal{J}'+\frac{k}{2\pi}\epsilon_\mathcal{J}'',\\
		\delta_\epsilon\tilde{\mathcal{L}}&=\epsilon_\mathcal{M}\tilde{\mathcal{M}}'+2\tilde{\mathcal{M}}\epsilon_\mathcal{M}'+\mathcal{P}\epsilon_\mathcal{P}'+\frac{k}{2\pi}\epsilon_\mathcal{P}''\nonumber\\
		&\quad+\epsilon_\mathcal{L}\tilde{\mathcal{L}}'+2\tilde{\mathcal{L}}\epsilon_\mathcal{L}'+\mathcal{J}\epsilon_\mathcal{J}',\\
		\delta_\epsilon\tilde{\mathcal{P}}&=\epsilon_\mathcal{L}\mathcal{P}'+\mathcal{P}\epsilon_\mathcal{L}'-\frac{k}{2\pi}\epsilon_\mathcal{J}'-\frac{k}{2\pi}\epsilon_\mathcal{L}'',\\
		\delta_\epsilon\tilde{\mathcal{J}}&=\epsilon_\mathcal{M}\mathcal{P}'+\mathcal{P}\epsilon_\mathcal{M}'-\frac{k}{2\pi}\epsilon_\mathcal{P}'-\frac{k}{2\pi}\epsilon_\mathcal{M}''\nonumber\\
		&\quad+\epsilon_\mathcal{L}\mathcal{J}'+\mathcal{J}\epsilon_\mathcal{L}'.
	\end{align}
	\end{subequations}
The reason why the redefinition \eqref{eq:RedefState} is beneficial is that it allows one to directly read off the Dirac bracket algebra of the functions $\tilde{\mathcal{M}}$, $\tilde{\mathcal{L}}$, $\mathcal{J}$ and $\mathcal{P}$ from \eqref{eq:GaugeTrafos} using e.g. 
	\begin{equation}
		\{\tilde{\mathcal{L}}(\varphi),\tilde{\mathcal{M}}(\bar{\varphi})\}=-\delta_{\epsilon_\mathcal{L}}\tilde{\mathcal{M}}(\bar{\varphi})
		\Bigr|_{\partial_{\bar{\varphi}}^n\epsilon_\mathcal{L}(\bar{\varphi})=(-1)^n\partial_\varphi^n\delta(\varphi-\bar{\varphi})}.
	\end{equation}
This trick works because the infinitesimal transformations \eqref{eq:GaugeTrafos} are related to the Dirac brackets of the canonical boundary charges as 
    \begin{equation}
        -\delta_{\epsilon_\mathcal{L}}\tilde{\mathcal{M}}(\bar{\varphi})=\{\mathcal{Q}[\epsilon_\mathcal{L}],\tilde{\mathcal{M}}(\bar{\varphi})\},
    \end{equation}
which reduces to
    \begin{equation}
        -\delta_{\epsilon_\mathcal{L}}\tilde{\mathcal{M}}(\bar{\varphi})=
		\int\extd\varphi\,\epsilon_\mathcal{L}(\varphi)\{\tilde{\mathcal{L}}(\varphi),\tilde{\mathcal{M}}(\bar{\varphi})\},
	\end{equation}
in case all coefficients in front of the canonical boundary charges are equal to one.

%---------------------------------------------------------------------------------------------------------------
\subsection{Canonical Boundary Charges}
%---------------------------------------------------------------------------------------------------------------

Having determined the boundary condition preserving gauge transformations \eqref{eq:BCPGTs} the next step is to determine the variation of the canonical boundary charges associated to these transformations. Using the Chern-Simons formalism these charges can be determined via\footnote{For more details see e.g. \cite{Henneaux:1992,Blagojevic:2002aa}.}
	\begin{equation}
		\delta Q[\epsilon] = \frac{k}{2\pi} \int\extd\varphi\, \langle \epsilon\,\delta \mathcal{A}_{\varphi} \rangle.
	\end{equation}
In terms of the redefined variables \eqref{eq:RedefState} and the boundary conditions preserving gauge transformations \eqref{eq:BCPGTs} the variation of the canonical boundary charges take the very simple form
	\begin{equation}
		\delta Q[\epsilon] =\int \extd\varphi\left( \delta\tilde{\mathcal{M}}\epsilon_\mathcal{M}+\delta\tilde{\mathcal{L}}\epsilon_\mathcal{L}+\delta\mathcal{P}\epsilon_\mathcal{P}+\delta\mathcal{J}\epsilon_\mathcal{J}\right).
	\end{equation}
This expression can easily be functionally integrated and leads to the following finite canonical boundary charge:
	\begin{equation}\label{eq:CSCanonicalBoundaryCharge}
		Q[\epsilon] =\int \extd\varphi\left(\tilde{\mathcal{M}}\epsilon_\mathcal{M}+\tilde{\mathcal{L}}\epsilon_\mathcal{L}+\mathcal{P}\epsilon_\mathcal{P}+\mathcal{J}\epsilon_\mathcal{J}\right).
	\end{equation}
One can readily check that this charge is also conserved in (retarded) time i.e. $\partial_u Q[\epsilon]=0$.

%---------------------------------------------------------------------------------------------------------------
\subsection{Asymptotic Symmetry Algebra}
%---------------------------------------------------------------------------------------------------------------

After having determined the canonical boundary charge \eqref{eq:CSCanonicalBoundaryCharge} one can -- with the help of \eqref{eq:GaugeTrafos} -- immediately determine the precise form of the asymptotic symmetry algebra. For the boundary conditions \eqref{eq:NewFSBCs} this yields the following non-vanishing Dirac brackets:
	\begin{subequations}
	\begin{align}
	\{\tilde{\mathcal{L}}(\varphi),\tilde{\mathcal{L}}(\bar{\varphi})\}&=2\tilde{\mathcal{L}}\delta'-\delta\tilde{\mathcal{L}}',\\
	\{\tilde{\mathcal{L}}(\varphi),\tilde{\mathcal{M}}(\bar{\varphi})\}&=2\tilde{\mathcal{M}}\delta'-\delta\tilde{\mathcal{M}}',\\
	\{\tilde{\mathcal{L}}(\varphi),\mathcal{J}(\bar{\varphi})\}&=\mathcal{J}\delta'-\delta\mathcal{J}',\\
	\{\tilde{\mathcal{L}}(\varphi),\mathcal{P}(\bar{\varphi})\}&=\mathcal{P}\delta'-\delta\mathcal{P}'-\frac{k}{2\pi}\delta'',\\
	\{\tilde{\mathcal{M}}(\varphi),\mathcal{J}(\bar{\varphi})\}&=\mathcal{P}\delta'-\delta\mathcal{P}'-\frac{k}{2\pi}\delta'',\\
	\{\mathcal{J}(\varphi),\mathcal{P}(\bar{\varphi})\}&=-\frac{k}{2\pi}\delta',
	\end{align}
	\end{subequations}
where all functions appearing on the r.h.s are functions of $\bar{\varphi}$ and prime denotes differentiation with respect to the corresponding argument. Moreover ${\delta\equiv\delta(\varphi-\bar{\varphi})}$ and $\delta'\equiv\partial_\varphi\delta(\varphi-\bar{\varphi})$. Expanding the fields and delta distribution in terms of Fourier modes as
	\begin{subequations}\label{eq:EinsteinFourierModes}
	\begin{align}
		\tilde{\mathcal{M}}&=\frac{1}{2\pi}\sum\limits_{n\in\mathbb{Z}}M_ne^{-in\varphi},&
		\tilde{\mathcal{L}}&=\frac{1}{2\pi}\sum\limits_{n\in\mathbb{Z}}L_ne^{-in\varphi},\\
		\mathcal{P}&=\frac{1}{2\pi}\sum\limits_{n\in\mathbb{Z}}P_ne^{-in\varphi},&
		\mathcal{J}&=\frac{1}{2\pi}\sum\limits_{n\in\mathbb{Z}}J_ne^{-in\varphi},\\
		\delta&=\frac{1}{2\pi}\sum\limits_{n\in\mathbb{Z}}e^{-in(\varphi-\bar{\varphi})},
	\end{align}
	\end{subequations}
and then replacing the Dirac brackets with commutators using $i\{\cdot,\cdot\}\rightarrow[\cdot,\cdot]$ one obtains the following non-vanishing commutation relations:
	\begin{subequations}\label{eq:ASAPrelim}
	\begin{align}
		[L_n,L_m]&=(n-m)L_{n+m},\\
		[L_n,M_m]&=(n-m)M_{n+m},\label{eq:BMSCentralTermMissing}\\
		[L_n,J_m]&=-mJ_{n+m},\\
		[L_n,P_m]&=-mP_{n+m}+ikn^2\delta_{n+m,0},\\
		[M_n,J_m]&=-mP_{n+m}+ikn^2\delta_{n+m,0},\\
		[J_n,P_m]&=-kn\delta_{n+m,0},
	\end{align}
	\end{subequations}
which is the asymptotic symmetry algebra of asymptotically flat spacetimes in 2+1 dimensions for the boundary conditions \eqref{eq:NewFSBCs}. One can already see at first glance some crucial differences to the $\mathfrak{bms}_3$ algebra found in \cite{Barnich:2006av}.\\
The first obvious differences are the additional (twisted) $\hat{\mathfrak{u}}(1)$ symmetries. The second important difference is the absence of a central term in \eqref{eq:BMSCentralTermMissing}. However, this is just because of our choice of basis. The usual $\mathfrak{bms}_3$ central charge can be recovered by a (twisted) Sugawara shift of $L_n$ and $M_n$ that takes the form
	\begin{equation}\label{SugShift}
		\hat{L}_n:=L_n+inJ_n\quad\hat{M}_n:=M_n+inP_n.
	\end{equation}
After this shift the algebra \eqref{eq:ASAPrelim} now reads
	\begin{subequations}\label{eq:ASAMiddle}
	\begin{align}
		[\hat{L}_n,\hat{L}_m]&=(n-m)\hat{L}_{n+m}\label{eq:VirasoroCentral},\\
		[\hat{L}_n,\hat{M}_m]&=(n-m)\hat{M}_{n+m}+\frac{c_M}{12}n^3\delta_{n+m,0},\\
		[\hat{L}_n,J_m]&=-mJ_{n+m},\\
		[\hat{L}_n,P_m]&=-mP_{n+m},\\
		[\hat{M}_n,J_m]&=-mP_{n+m},\\
		[J_n,P_m]&=\kappa_P n\,\delta_{n+m,0},\label{eq:AffU1}
	\end{align}
	\end{subequations}
with $c_M=12k$ and $\kappa_P=-k$. Since $k=\frac{1}{4G_N}$ one has thus successfully recovered the $\mathfrak{bms}_3$ central charge.\\
In order to make the structure of the two additional affine $\hat{\mathfrak{u}}(1)$ current algebras that are contained within \eqref{eq:AffU1} more apparent one can also make the following redefinition:
	\begin{equation}
		\hat{J}_n^\pm:=\frac{1}{\sqrt{2}}\left(J_n\pm P_n\right).
	\end{equation}
With this redefinition the commutator \eqref{eq:AffU1} changes into
	\begin{equation}
		[\hat{J}^\pm_n,\hat{J}^\pm_m]=\kappa_P n\delta_{n+m,0}.
	\end{equation}
Thus the boundary conditions \eqref{eq:NewFSBCs} can be seen as an extension of the existing Barnich-Comp{\`e}re boundary conditions by two affine $\hat{\mathfrak{u}}(1)$ current algebras with level $\kappa_P=c_M=\frac{1}{4G_N}$.

%---------------------------------------------------------------------------------------------------------------
\section{Extension to Topologically Massive Gravity}\label{sec:TMG}
%---------------------------------------------------------------------------------------------------------------

In this section we extend our findings to TMG, described by the action \cite{Deser:1981wh,Deser:1982vy}
	\begin{equation}\label{eq:TMGAction}
		I_{\textnormal{TMG}}=\frac{1}{16\pi G_N}\int\extd^3x\sqrt{-g}\left(R+\frac{1}{2\mu}CS[\Gamma]\right),
	\end{equation}
where $CS[\Gamma]=\varepsilon^{\lambda\mu\nu}\Gamma^\rho{}_{\lambda\sigma}\left(\partial_\mu\Gamma^\sigma{}_{\rho\nu}+\frac{2}{3}\Gamma^\sigma{}_{\mu\tau}\Gamma^\tau{}_{\nu\rho}\right)$ is a gravitational Chern-Simons term and $\mu$ is the corresponding Chern-Simons coupling. In order to proceed we first translate the results from Section \ref{sec:CS} to a metric formulation.

%---------------------------------------------------------------------------------------------------------------
\subsection{Boundary Conditions}
%---------------------------------------------------------------------------------------------------------------

The boundary conditions found in the Chern-Simons formalism can be translated to the metric formulation by extracting the dreibein $e$ from the Chern-Simons connection $\mathcal{A}$ via
	\begin{equation}
		\mathcal{A}=\omega^aL_a+e^aM_a,
	\end{equation}
for $a=0,\pm1$. Then using
	\begin{equation}
		\eta_{ab}=-2\left(
			\begin{array}{c|ccc}
				  &M_1&M_0&M_{-1}\\
				\hline
				M_1&0&0&1\\
				M_0&0&-\frac{1}{2}&0\\
				M_{-1}&1&0&0
			\end{array}\right),
	\end{equation}
one can recover a metric formulation via 
	\begin{equation}
		g_{\mu\nu}=\eta_{ab}e^a_\mu e^b_\nu.
	\end{equation}
For the boundary conditions \eqref{eq:NewFSBCs} this leads to the following metric:
	\begin{align}\label{eq:AdSBMSMetricNew}
		\extd s^2&=\mathcal{M}\extd u^2+2\mathcal{N}\extd u\extd\varphi-2e^{\alpha}\extd r\extd u\nonumber\\
		&\quad+\left[e^{2\alpha}r^2+\beta\left(2\mathcal{N}-\mathcal{M}\beta\right)\right]\extd\varphi^2.
	\end{align}
Starting from \eqref{eq:AdSBMSMetricNew} one can infer that the following fluctuations of the metric can be allowed:
	\begin{subequations}\label{eq:BCsMetric}
	\begin{align}
		g_{rr}&=\mathcal{O}(r^{-2}),\\
		g_{r\varphi}&=-e^\alpha\beta+\mathcal{O}(r^{-1}),\\
		g_{ru}&=-e^\alpha+\mathcal{O}(r^{-1}),\\
		g_{\varphi\varphi}&=e^{2\alpha}r^2+\beta\left(2\mathcal{N}-\mathcal{M}\beta\right)+\mathcal{O}(r^{-2}),\\
		g_{\varphi u}&=\mathcal{N}+\mathcal{O}(r^{-1}),\\
		g_{uu}&=\mathcal{M}+\mathcal{O}(r^{-1}).
	\end{align}
	\end{subequations}
The procedure to determine the asymptotic symmetries is in principle the same as in the Chern-Simons formulation. First one has to find the asymptotic Killing vectors that preserve the asymptotic structure of \eqref{eq:BCsMetric}. Then using these Killing vectors one can compute the associated canonical charges as well as the algebra these Killing vectors have to satisfy.

%---------------------------------------------------------------------------------------------------------------
%\subsection{Variational Principle}
%---------------------------------------------------------------------------------------------------------------

%In order to be able to check the variational principle one first has to change from the Eddington-Finkelstein coordinates that have been employed so far to a FSC-like coordinate system. One possible way to achieve this is given by the following change of coordinates:
%    \begin{subequations}
%    \begin{align}
%        \rho&=re^\alpha,\\
%        \extd t&=\extd u+\beta\extd\varphi-\frac{\rho^2\extd\rho}{\rho_+^2(\rho^2-\rho_0^2)},\\
%        \extd \phi&=\extd\varphi-\frac{\rho_0\extd\rho}{\rho_+(\rho^2-\rho_0^2)},
%    \end{align}
%    \end{subequations}
%with
%    \begin{equation}
%        \rho_+=\sqrt{\mathcal{M}},\qquad\rho_0=\frac{\beta\mathcal{M}-\mathcal{N}}{\sqrt{\mathcal{M}}},
%    \end{equation}
%and where all functions $\mathcal{M}$, $\mathcal{N}$, $\alpha$ and $\beta$ are assumed to be constant.\\    
%In terms of the new coordinates $\rho$, $t$ and $\phi$ the metric then reads
%    \begin{equation}
%        \extd s^2=\rho_+^2\extd t^2-\frac{\rho^2\extd\rho^2}{\rho_+^2(\rho^2-\rho_0^2)}+\rho^2\extd\phi^2-2\rho_+\rho_0\extd t\extd\phi.
%   \end{equation}
%---------------------------------------------------------------------------------------------------------------
\subsection{Asymptotic Killing Vectors}
%---------------------------------------------------------------------------------------------------------------

The Killing vectors that preserve the asymptotic structure of \eqref{eq:BCsMetric} are given by the vector fields $\xi^\mu$ satisfying
	\begin{subequations}
	\begin{align}
		\mathcal{L}_\xi g_{rr}&=\mathcal{O}(r^{-2}),\\
		\mathcal{L}_\xi g_{r\varphi}&=-\delta\left(e^\alpha\beta\right)+\mathcal{O}(r^{-1}),\\
		\mathcal{L}_\xi g_{ru}&=\delta e^\alpha+\mathcal{O}(r^{-1}),\\
		\mathcal{L}_\xi g_{\varphi\varphi}&=\delta e^{2\alpha}r^2+\delta\left(\beta\left(2\mathcal{N}-\mathcal{M}\beta\right)\right)+\mathcal{O}(r^{-2}),\\
		\mathcal{L}_\xi g_{rr}&=\delta\mathcal{N}+\mathcal{O}(r^{-1}),\\
		\mathcal{L}_\xi g_{r\varphi}&=\delta\mathcal{M}+\mathcal{O}(r^{-1}),
	\end{align}
	\end{subequations}
where the $\delta$ symbolizes the infinitesimal change of the corresponding functions under the action of the asymptotic Killing vector $\xi$. The Killing vectors satisfying these conditions are given by
	\begin{subequations}\label{eq:AKV}
	\begin{align}
		\xi^r&=\mathfrak{J}r+e^{-\alpha}(\beta\mathfrak{J}'-\mathfrak{P}')+\mathcal{O}(r^{-1}),\\
		\xi^\varphi&=\mathfrak{L}+\frac{e^{-\alpha}\mathfrak{P}}{r}+\mathcal{O}(r^{-2}),\\
		\xi^u&=\mathfrak{M}-\frac{e^{-\alpha}\beta\mathfrak{P}}{r}+\mathcal{O}(r^{-2}),
	\end{align}
	\end{subequations}
where  $\mathfrak{L}$, $\mathfrak{M}$, $\mathfrak{J}$ and $\mathfrak{P}$ are %in principle arbitrary 
functions of $u$ and $\varphi$ that have to satisfy
	\begin{equation}\label{eq:KillingTimeEvolution}
		\partial_u\mathfrak{L}=\partial_u\mathfrak{J}=0,\,\partial_u\mathfrak{M}=\partial_\varphi\mathfrak{L},\,\partial_u\mathfrak{P}=\partial_\varphi\mathfrak{J}.
	\end{equation}	
Hence, from these equations we have that 
\begin{eqnarray}
\mathfrak{J} &=& \mathfrak{j}(\varphi) \\
\mathfrak{L} &=& \mathfrak{l}(\varphi)\\
\mathfrak{P} &=& \mathfrak{p}(\varphi) + u \mathfrak{j}' \\
\mathfrak{M} &=& \mathfrak{m}(\varphi) + u \mathfrak{l}'.
\end{eqnarray}

%---------------------------------------------------------------------------------------------------------------
\subsection{Killing Vector Algebra}
%---------------------------------------------------------------------------------------------------------------

After having found the asymptotic Killing vectors the next step is to determine the algebra these vector satisfy. Since the Killing vectors \eqref{eq:AKV} are state dependent one has to use modified \cite{Barnich:2010eb} (or ``adjusted'' \cite{Compere:2015knw}) Lie brackets that are defined as
	\begin{equation}\label{eq:ModifiedBracket}
		[\xi_1,\xi_2]^\mu_*=[\xi_1,\xi_2]-\delta_{\xi_1}\xi_2+\delta_{\xi_2}\xi_1,
	\end{equation}
where $\delta_{\xi_1}\xi_2$ denotes the change of the vector field $\xi_2$ under an infinitesimal flow along the vector field $\xi_1$ i.e. it takes into account the change the functions $\alpha$, $\beta$, $\mathcal{N}$ and $\mathcal{M}$ that appear both in the metric and the AKV $\xi_2$ undergo under a flow along $\xi_1$. This means e.g.
    \begin{equation}
        \delta_{\xi_1}\xi_2^\varphi=\frac{e^\alpha\delta_{\xi_1}\alpha\mathfrak{P}}{r},
    \end{equation}
where $\delta_{\xi_1}\alpha$ denotes the infinitesimal change $\alpha$ undergoes upon acting on the metric with $\xi_1$ via $\mathcal{L}_{\xi_1}g_{\mu\nu}$.\\
For the Killing vectors \eqref{eq:AKV} one can write the evaluation of the bracket \eqref{eq:ModifiedBracket} in a compact form as
	\begin{equation}
		[\xi_1,\xi_2]^\mu_*=\xi_{[1,2]}^\mu,
	\end{equation}
where for the functions $\mathfrak{l}$, $\mathfrak{m}$, $\mathfrak{j}$ and $\mathfrak{p}$ appearing in the Killing vectors one finds
	\begin{subequations}
	\begin{align}
		\mathfrak{l}_{[1,2]}&=\mathfrak{l}_1\mathfrak{l}_2'-\mathfrak{l}_2\mathfrak{l}_1',\\
		\mathfrak{m}_{[1,2]}&=\mathfrak{m}_1\mathfrak{l}_2'-\mathfrak{m}_2\mathfrak{l}_1'+\mathfrak{l}_1\mathfrak{m}_2'-\mathfrak{l}_2\mathfrak{m}_1',\\
		\mathfrak{j}_{[1,2]}&=\mathfrak{l}_1\mathfrak{j}_2'-\mathfrak{l}_2\mathfrak{j}_1',\\
		\mathfrak{p}_{[1,2]}&=\mathfrak{m}_1\mathfrak{j}_2'-\mathfrak{m}_2\mathfrak{j}_1'+\mathfrak{l}_1\mathfrak{p}_2'-\mathfrak{l}_2\mathfrak{p}_1',
	\end{align}
	\end{subequations}
and a prime denotes a derivative with respect to $\varphi$.\\
For a Killing vector
%$\xi^\mu(\mathfrak{L},\mathfrak{M},\mathfrak{J},\mathfrak{P})$
$\xi^\mu(\mathfrak{l},\mathfrak{m},\mathfrak{j},\mathfrak{p})$ one can introduce Fourier modes as
	\begin{subequations}\label{eq:FourierModes}
	\begin{align}
		L_n&=\xi^\mu(e^{in\varphi},0,0,0),&M_n&=\xi^\mu(0,e^{in\varphi},0,0),\\
		J_n&=\xi^\mu(0,0,e^{in\varphi},0),&P_n&=\xi^\mu(0,0,0,e^{in\varphi}).
	\end{align}
	\end{subequations}
In terms of these Fourier modes the algebra of the asymptotic Killing vectors looks like
	\begin{subequations}
	\begin{align}
		i[L_n,L_m]&=(n-m)L_{n+m},\\
		i[L_n,M_m]&=(n-m)M_{n+m},\\
		i[L_n,J_m]&=-mJ_{n+m},\\
		i[L_n,P_m]&=-mP_{n+m},\\
		i[M_n,J_m]&=-mP_{n+m},
	\end{align}
	\end{subequations}
which is basically the algebra \eqref{eq:ASAMiddle} without central extensions. This is a nice check that, indeed, or boundary conditions also translate nicely to a metric formulation. The next step is again to determine the canonical boundary charge in order to see how the addition of the topological Chern-Simons term in \eqref{eq:TMGAction} changes the boundary dynamics in comparison to the pure Einstein gravity case considered in Section \ref{sec:CS}.

%---------------------------------------------------------------------------------------------------------------
\subsection{Canonical Boundary Charges}
%---------------------------------------------------------------------------------------------------------------

The conserved charges in TMG are written as \cite{Abbott:1981ff,Iyer:1994ys,Deser:2003vh,Bouchareb:2007yx,Compere:2008cv}
\begin{equation}\label{eq:chargevar}
\delta Q_\xi= \frac{1}{16\pi G_N}\int k_\xi[g, h] \ ,
\end{equation}
where $\xi$ is an asymptotic Killing vector, and $h_{\mu\nu} = \delta g_{\mu\nu}$ a linearized solution to the equations of motion. Expression \eqref{eq:chargevar} then gives the infinitesimal charge difference between two solutions $g$ and $g+\delta g$ to the equations of motion. The integrand appearing in \eqref{eq:chargevar} is written as \begin{equation}\label{k}
k_\xi[g, h] = \epsilon_{\mu\nu\rho}\left(k_{grav}^{\mu\nu}[\zeta; h,g] + k_{cs}^{\mu\nu}[\zeta; h,g] \right)dx^\rho \ .
\end{equation}
The Einstein contribution to this expression is
\begin{align}
k_{grav}^{\mu\nu}[\zeta; h, g] = & \zeta^\nu(D^\mu h - D_\sigma h^{\mu\sigma}) + \zeta_\sigma D^\nu h^{\mu\sigma}\nonumber \\
& + \frac{1}{2} h D^\nu \zeta^\mu - h^{\nu\sigma}D_\sigma\zeta^\mu\nonumber\\
& + \frac{1}{2}h^{\sigma \nu}(D^\mu\zeta_\sigma + D_\sigma \zeta^\mu) ,
\end{align}
and the Chern-Simons contribution
\begin{align}
k_{cs}^{\mu\nu}[\zeta; h, g] = & \frac{1}{\mu}k_{grav}^{\mu\nu}[\eta; h, g] \nonumber\\ 
& +\frac{1}{2\mu}\epsilon^{\mu\nu\rho}\left(\zeta_\rho h^{\lambda \sigma}G_{\sigma\lambda} + \frac{1}{2}h(\zeta_\sigma G^\sigma_{\ \rho} + \frac{1}{2}\zeta_\rho R)\right)\nonumber\\
& - \frac{1}{2\mu}\zeta_\lambda\left(2 \epsilon^{\mu\nu\rho}\delta(G^\lambda_{\ \rho}) - \epsilon^{\mu\nu\lambda}\delta G\right),
\end{align}
where $\eta^\mu = \frac{1}{2}\epsilon^{\mu\nu\rho}D_\nu\zeta_\rho$. 
Finite charges are obtained by integrating the variation \eqref{eq:chargevar} from one solution to another.\\
The charges associated with the AKV (\ref{eq:AKV}) can be computed for the metric (\ref{eq:AdSBMSMetricNew}) using the on-shell relations \eqref{eq:StateOnShell}.
%On-shell, the functions appearing in the metric take the following form:
%\begin{subequations}
%\begin{align}
%\mathcal{M} &=&  M(\varphi) \\
%\alpha &=& A(\varphi)\\
%\mathcal{N} &=& N(\varphi) +\frac{u}{2} M' \\
%\beta &=& B(\varphi) + u A'.
%\end{align}
%\end{subequations}
The total integrated charge can then be written as
\begin{equation}
Q_\xi = Q_\mathfrak{l} + Q_\mathfrak{m} + Q_\mathfrak{j}+ Q_\mathfrak{p}, 
\end{equation}
with
\begin{subequations}\label{eq:TMGCharges}
\begin{align}
Q_\mathfrak{l} &= \frac{1}{8 \pi G_N} \int \mathfrak{l} \mathcal{L} -
    \frac{1}{4 \mu}\mathfrak{l} (2 \mathcal{M} - A'^2 + 2 A''),\\
Q_\mathfrak{m} &= \frac{1}{16 \pi G_N} \int \mathfrak{m} \mathcal{M}, 
\\
Q_\mathfrak{j} &= -\frac{1}{8 \pi G_N} \int \mathfrak{j} B' +
    \frac{3}{2\mu}\mathfrak{j} A', 
\\
Q_\mathfrak{p} &= -\frac{1}{8 \pi G_N} \int \mathfrak{p} A'.
\end{align}
\end{subequations}
In order to determine the asymptotic symmetry algebra associated with these charges it will again prove useful to introduce new functions as
	\begin{subequations}\label{eq:RedefStateTMG}
	\begin{align}
		\tilde{\mathcal{M}}&=\frac{\mathcal{M}}{16\pi G_N},\quad\mathcal{P}=-\frac{A'}{8\pi G_N},\\
		\tilde{\mathcal{L}}&=\frac{\mathcal{L}}{8\pi G_N}-\frac{1}{4\mu}\left(4\tilde{\mathcal{M}}-\frac{2\pi}{k}\mathcal{P}^2-2\mathcal{P}'\right),\\
		\mathcal{J}&=-\frac{B'}{8\pi G_N}+\frac{3}{2\mu}\mathcal{P}.
	\end{align}
	\end{subequations}    
Using these redefinitions one can write the charges in a more compact way as
    \begin{equation}
        Q_\xi=\int\mathfrak{l}\tilde{\mathcal{L}}+\mathfrak{m}\tilde{\mathcal{M}}+\mathfrak{j}\mathcal{J}+\mathfrak{p}\mathcal{P},
    \end{equation}
which is again the general form as in the Einstein case \eqref{eq:CSCanonicalBoundaryCharge}. It is important to note, however, that the addition of the gravitational Chern-Simons term deformed some of the state dependent functions and thus will also deform the central extensions.

%---------------------------------------------------------------------------------------------------------------
\subsection{Asymptotic Symmetry Algebra}
%---------------------------------------------------------------------------------------------------------------

Using the same trick as in the Einstein case i.e. reading off the Dirac brackets from the infinitesimal transformations of the state dependent functions one finds the following algebra for TMG:
	\begin{subequations}
	\begin{align}
	\{\tilde{\mathcal{L}}(\varphi),\tilde{\mathcal{L}}(\bar{\varphi})\}&=2\tilde{\mathcal{L}}\delta'-\delta\tilde{\mathcal{L}}'-\frac{k}{4\pi\mu}\delta''',\\
	\{\tilde{\mathcal{L}}(\varphi),\tilde{\mathcal{M}}(\bar{\varphi})\}&=2\tilde{\mathcal{M}}\delta'-\delta\tilde{\mathcal{M}}',\\
	\{\tilde{\mathcal{L}}(\varphi),\mathcal{J}(\bar{\varphi})\}&=\mathcal{J}\delta'-\delta\mathcal{J}'+\frac{3k}{4\pi\mu}\delta'',\\
	\{\tilde{\mathcal{L}}(\varphi),\mathcal{P}(\bar{\varphi})\}&=\mathcal{P}\delta'-\delta\mathcal{P}'+\frac{k}{2\pi}\delta'',\\
	\{\tilde{\mathcal{M}}(\varphi),\mathcal{J}(\bar{\varphi})\}&=\mathcal{P}\delta'-\delta\mathcal{P}'+\frac{k}{2\pi}\delta'',\\
	\{\mathcal{J}(\varphi),\mathcal{J}(\bar{\varphi})\}&=-\frac{3k}{4\pi\mu}\delta',\\
	\{\mathcal{J}(\varphi),\mathcal{P}(\bar{\varphi})\}&=-\frac{k}{2\pi}\delta'.
	\end{align}
	\end{subequations}
In order to evaluate the effect of the gravitational Chern-Simons term on the asymptotic symmetry algebra, one first has to perform basically the same shift as in the Einstein gravity case as
    \begin{equation}
        \hat{\mathcal{L}}=\tilde{\mathcal{L}}+\mathcal{J}',\quad\hat{\mathcal{M}}=\tilde{\mathcal{M}}+\mathcal{P}'.
    \end{equation}
Introducing Fourier modes as in \eqref{eq:EinsteinFourierModes} one then obtains the following non-vanishing commutation relations:
	\begin{subequations}\label{eq:ASATMG}
	\begin{align}
		[\hat{L}_n,\hat{L}_m]&=(n-m)\hat{L}_{n+m}+\frac{c_L}{12}n^3\delta_{n+m,0},\\
		[\hat{L}_n,\hat{M}_m]&=(n-m)\hat{M}_{n+m}+\frac{c_M}{12}n^3\delta_{n+m,0},\\
		[\hat{L}_n,J_m]&=-mJ_{n+m},\\
		[\hat{L}_n,P_m]&=-mP_{n+m},\\
		[\hat{M}_n,J_m]&=-mP_{n+m},\\
		[J_n,J_m]&=\kappa_J\,n\delta_{n+m,0},\\
		[J_n,P_m]&=\kappa_P\,n\delta_{n+m,0},
	\end{align}
	\end{subequations}
where $c_M$ and $\kappa_P$ are the same as in the Einstein gravity case and	$c_L=\frac{3}{\mu G_N}$ and $\kappa_J=-\frac{3}{8\mu G_N}$.\\
Finally, it is noteworthy that in the Flat Space Chiral Gravity limit \cite{Bagchi:2012yk}, $G_N \rightarrow \infty$, $\mu G_N$ fixed, we are left with a chiral Virasoro algebra supplemented by the affine $\hat{\mathfrak{u}}(1)$ current algebra generated by $J_n$.

%---------------------------------------------------------------------------------------------------------------
\section{Thermal Entropy}\label{sec:ThermalEntropy}
%---------------------------------------------------------------------------------------------------------------

Having determined the asymptotic symmetries the next interesting question to ask is: How does the thermal entropy of solutions that obey \eqref{eq:NewFSBCs} look like both in Einstein gravity and TMG and what is the geometric interpretation? In the following two subsections we first give a possible geometric interpretation of our solutions by locating the cosmological horizon and then determining the associated thermal entropy using a metric formulation of Einstein gravity. We then present a second, independent derivation of the thermal entropy in Einstein gravity using the Chern-Simons formulation. In the last part of this section we then determine the Chern-Simons contribution to the Bekenstein-Hawking entropy determined previously in the case of TMG by integrating the first law of flat space cosmologies.

%--------------------------------------------------
\subsection{Horizon Area}
%---------------------------------------------------------------------------------------------------------------

In the metric formalism the relevant quantity to compute the thermal entropy is given by the horizon area. In order to compute the area of the horizon one first has to locate the event horizon by looking at the point where the determinant of the induced metric on slices of constant radius vanishes, that is
    \begin{equation}
        g_{uu}\,g_{\varphi\varphi}-\left(g_{u\varphi}\right)^2=0.
    \end{equation}
For the metric \eqref{eq:AdSBMSMetricNew} this happens at
    \begin{equation}\label{eq:Horizon}
        r_{\rm H}=\frac{|\mathcal{N}-\mathcal{M}\beta|}{\sqrt{\mathcal{M}}}e^{-\alpha}.
    \end{equation}
One can now compute the area of the horizon by
    \begin{equation}
        {\rm A}_{\rm H}=\int\limits_0^{2\pi}\extd\varphi\sqrt{|g_{\varphi\varphi}|}\,\Big|_{r=r_{\rm H}},
    \end{equation}
that for the case at hand reduces to
    \begin{equation}\label{eq:HorizonArea}
        {\rm A}_{\rm H}=\int\limits_0^{2\pi}\extd\varphi\frac{\mathcal{N}}{\sqrt{\mathcal{M}}}.
    \end{equation}
For the zero mode solutions $\mathcal{N}(\varphi)=\mathcal{L}=\,$const. and $\mathcal{M}(\varphi)=\mathcal{M}=\,$const., this can be trivially integrated to yield the area
    \begin{equation}
        {\rm A}_{\rm H}^0=2\pi\frac{\mathcal{L}}{\sqrt{\mathcal{M}}},
    \end{equation}
and the corresponding thermal entropy
    \begin{equation}\label{eq:ThermalEntropyArea}
        S_{\rm Th}=\frac{{\rm A}_{\rm H}^0}{4G_N}=\frac{\pi}{2G_N}\frac{\mathcal{L}}{\sqrt{\mathcal{M}}},
    \end{equation}
which is the thermal entropy one would expect from an ordinary FSC \cite{Bagchi:2012xr,Barnich:2012xq,Bagchi:2013qva}.\\
This result is rather surprising since the asymptotic symmetry algebra contains two spin-1 currents in addition to the usual mass and angular momentum of a FSC and one might expect contributions of these currents to the thermal entropy.\\
Looking at \eqref{eq:Horizon} one can see that for general, $\varphi$ dependent $\alpha$ and $\beta$, these two functions change the location of the horizon radius in an angle dependent way and thus describe non-spherically symmetric cosmological solutions\footnote{It should be noted that also the authors of \cite{Afshar:2016kjj} found non-spherically symmetric solutions in the context of flat space Einstein gravity that they called "cosmological flowers". While both the solutions presented in this paper and the ones in \cite{Afshar:2016kjj} describe non-spherically symmetric cosmological solutions the associated canonical charges are different. Also in the context of New Massive Gravity (NMG) \cite{Bergshoeff:2009hq} non-spherically symmetric cosmological solutions with different canonical charges have been already described in \cite{Barnich:2015dvt}.}. However, they do it in such a way that the area of the cosmological solutions remains the same as for constant values of $\alpha$ and $\beta$.

%--------------------------------------------------
\subsection{First Law, Holonomy Conditions and Wilson Lines}
%---------------------------------------------------------------------------------------------------------------

In this section we show how the validity of a first law for FSCs can be used to determine the thermal entropy in the Chern-Simons formulation if the holonomies of the connection $\mathcal{A}$ satisfy certain requirements in close analogy to the derivations found in \cite{Gutperle:2011kf,Ammon:2011nk,Perez:2012cf,Perez:2013xi,deBoer:2013gz} for the AdS$_3$ (higher-spin) case. Furthermore we show that this derivation is equivalent to a Wilson line wrapping around the non-contractible $\varphi$ cycle.\\
FSCs with inverse temperature $\beta_T$, mass $M$, angular velocity $\Omega$ and angular momentum $J$ satisfy the first law of flat space cosmologies \cite{Bagchi:2012xr} that is given by
    \begin{equation}\label{eq:FirstLaw}
        \delta M=-T\delta S_{\textrm{Th}}+\Omega\delta J.
    \end{equation}
In a metric formulation the mass and angular momentum are associated to the charges of the Killing vectors $\partial_u$ and $\partial_\varphi$ respectively. Using that the gauge parameters \eqref{eq:GaugeParameter} that preserve the boundary conditions \eqref{eq:NewFSBCs} are related to the AKVs via $\epsilon=\xi^\mu\mathcal{A}_\mu$ one can determine (the variation of) mass and angular momentum of our cosmological solutions via
    \begin{subequations}\label{eq:VariationMassAngMomFormula}
    \begin{align}
        \delta M &:= \delta Q[\partial_u]=\frac{k}{2\pi}\int\extd\varphi\langle\mathcal{A}_u\delta\mathcal{A}_\varphi\rangle,\\
        \delta J &:= \delta Q[\partial_\varphi]=\frac{k}{2\pi}\int\extd\varphi\langle\mathcal{A}_\varphi\delta\mathcal{A}_\varphi\rangle.
    \end{align}
    \end{subequations}
Plugging in our expression for the connection \eqref{eq:NewFSBCs} one sees that only the zero modes of $\mathcal{N}(\varphi)$ and $\mathcal{M}(\varphi)$ that we again denoted by $\mathcal{L}$ and $\mathcal{M}$ respectively contribute in \eqref{eq:VariationMassAngMomFormula} as
    \begin{equation}
        \delta M = \frac{k}{2}\delta\mathcal{M},\qquad\delta J = k \delta \mathcal{L}.
    \end{equation}
So already at this level it is apparent that the two additional $\hat{\mathfrak{u}}(1)$ currents do not have any influence on the mass or angular momentum of our cosmological solutions.\\
After having determined mass and angular momentum the remaining pieces of the puzzle to determine the entropy via the first law \eqref{eq:FirstLaw} is to determine the inverse temperature $\beta_T$ and angular velocity $\Omega$. We will do so by imposing conditions on the holonomy $e^{ih}$ with
    \begin{equation}\label{eq:HolonomyDefinition}
        h=-\frac{\beta_T}{2\pi}\left(\int\extd\varphi\,a_u - \Omega \int\extd\varphi\,a_\varphi\right).
    \end{equation}
Before stating what these conditions are it may be illuminating to rewrite the first law \eqref{eq:FirstLaw} using \eqref{eq:VariationMassAngMomFormula}
    \begin{equation}\label{eq:VariationEntropy}
        \delta S_{\textrm{Th}}=-\frac{k}{2\pi}\beta_T\int\extd\varphi \langle a_u\delta a_\varphi\rangle+\frac{k}{2\pi}\beta_T\,\Omega\int\extd\varphi \langle a_\varphi\delta a_\varphi\rangle,
    \end{equation}
and then using \eqref{eq:HolonomyDefinition} to yield
    \begin{equation}\label{eq:EntropyAndHol}
        \delta S_{\textrm{Th}} = k\langle h\,\delta a_\varphi\rangle.
    \end{equation}
We want to point out that aside from assuming that the solution in question satisfies a first law we did not make any additional assumptions. Thus \eqref{eq:EntropyAndHol} holds in particular also for BTZ black holes\footnote{Modulo the different sign of the temperature that can be taken care of by exchanging $h\rightarrow-h$.}. Looking at the BTZ case one finds that the eigenvalues of $h$ conspire in such a way that
    \begin{equation}\label{eq:HolCond}
        \textrm{Eigen}\left[h\right]= \textrm{Eigen}\left[2\pi L_0\right].
    \end{equation}
That means in particular that the holonomy associated with $h$ is trivial\footnote{To be more precise, one has $e^{ih}=-\unity$.}. In addition this also means that for the case where \eqref{eq:HolCond} is satisfied one can also functionally integrate \eqref{eq:EntropyAndHol} to yield
    \begin{equation}\label{eq:EntropyAndHol2}
        S_{\textrm{Th}} = k\langle h\,a_\varphi\rangle.
    \end{equation}    
In close analogy to the BTZ case we are also assuming that \eqref{eq:HolCond} holds for our cosmological solutions. This assumption yields the following restrictions on the inverse temperature $\beta_T$ and angular velocity $\Omega$ after integrating over $\varphi$ in \eqref{eq:HolonomyDefinition}
    \begin{equation}\label{eq:InvTempAndAngVel}
        \beta_T=\frac{2\pi\mathcal{L}}{\mathcal{M}^{\frac{3}{2}}},\qquad \Omega=\frac{\mathcal{L}}{\mathcal{M}}.
    \end{equation}
Now using \eqref{eq:EntropyAndHol2} one immediately finds the thermal entropy to be
    \begin{equation}
        S_{\textrm{Th}} = \frac{2\pi k \mathcal{L}}{\mathcal{M}} = \frac{\pi \mathcal{L}}{2G_N\mathcal{M}},
    \end{equation}
which is exactly the same result \eqref{eq:ThermalEntropyArea} that has been found in the previous section via the area law.\\
Another advantage of writing the entropy as in \eqref{eq:EntropyAndHol2} is that the relation to a Wilson line wrapping around the horizon\footnote{How to employ Wilson lines to determine both entanglement entropy and thermal entropy of BTZ black holes has been first shown in \cite{Ammon:2013hba,deBoer:2013vca}. This approach was then later modified for applications in flat space in \cite{Bagchi:2014iea,Basu:2015evh,Riegler:2016hah}.} is almost apparent if the holonomy conditions \eqref{eq:HolCond} are satisfied. Using this approach one can write the thermal entropy in a very compact way
    \begin{equation}\label{eq:EntropyWilsonLine}
        S_{\rm Th}=\frac{\pi}{6}c_M\langle L_0\lambda_\varphi\rangle,
    \end{equation}
where $\lambda_\varphi$ denotes the diagonalized version of $a_\varphi$. For $h=2\pi L_0$ one can see that \eqref{eq:EntropyAndHol2} and \eqref{eq:EntropyWilsonLine} coincide since the bilinear form yields the sum of the products of the eigenvalues of $L_0$ and $a_\varphi$. Thus \eqref{eq:EntropyAndHol2} is equivalent to a Wilson line wrapping around the horizon of a cosmological solution in flat space \cite{Bagchi:2014iea,Basu:2015evh,Riegler:2016hah} if the holonomy conditions \eqref{eq:HolCond} are satisfied.

%--------------------------------------------------
\subsection{The TMG Contribution to Thermal Entropy}
%--------------------------------------------------

In the case of TMG the thermal entropy contains in addition to the Bekenstein-Hawking piece a contribution proportional to the Chern-Simons coupling $\mu$. Thus the total thermal entropy can be written as
    \begin{equation}
        S^{\textrm{TMG}}_{\textrm{Th}}=\frac{{\rm A}_{\rm H}^0}{4G_N}+\frac{1}{\mu}S_{\rm{CS}}.
    \end{equation}
The Chern-Simons contribution $S_{\rm{CS}}$ can be determined in various ways (see e.g. \cite{Solodukhin:2005ah,Park:2006gt,Park:2006zw,Tachikawa:2006sz}). 
Since the first law is just a consequence of the conservation of $k_\xi[g, h]$ given in (\ref{k}), we can also simply integrate it, knowing the charges and the potentials.
The inverse temperature $\beta_T$ and the angular velocity $\Omega$ being theory independent quantities, one can use the expressions determined previously in \eqref{eq:InvTempAndAngVel} also for the TMG case. The definition of mass and angular momentum, however, are theory dependent and thus are different for TMG. In order to determine mass and angular momentum in TMG using our boundary conditions one can apply the same logic as in the previous section i.e. by looking at the charges \eqref{eq:TMGCharges} associated to the AKVs $\xi^u$ and $\xi^\varphi$ respectively. Focusing on solutions where $\mathcal{M}(\varphi)=\mathcal{M}$, $\mathcal{L}(\varphi)=\mathcal{L}$ and $A(\varphi)=A$ are constant one finds the following expressions for mass and angular momentum in terms of the corresponding quantities that appear in the metric:
    \begin{equation}
        M = \frac{\mathcal{M}}{16\pi G_N},\quad
        J = \frac{1}{8\pi G_N}\left(\mathcal{L}-\frac{\mathcal{M}}{2\mu}\right)
    \end{equation}
Using these expressions for mass and angular momentum one can write the first law as
    \begin{equation}
        \delta S_{\rm{Th}}^{\rm{TMG}}=\delta S_{\rm{Th}}^{\rm{EH}}+\frac{1}{\mu}\delta S_{\rm{CS}},
    \end{equation}
where $S_{\rm{Th}}^{\rm{EH}}$ is the thermal entropy given by \eqref{eq:ThermalEntropyArea} and
    \begin{equation}
        \delta S_{\rm{CS}}=-\frac{\pi\delta\mathcal{M}}{4G_N\sqrt{\mathcal{M}}}.
    \end{equation}
This expression for $\delta S_{\rm{CS}}$ can be easily functionally integrated to yield the total thermal entropy
    \begin{equation}\label{eq:TMGThermalEntropy}
        S^{\textrm{TMG}}_{\textrm{Th}}=\frac{\pi}{2G_N}\left(\frac{\mathcal{L}}{\sqrt{\mathcal{M}}}-\frac{\sqrt{M}}{\mu}\right),
    \end{equation}
that again perfectly agrees with the results obtained in \cite{Bagchi:2013lma} for spherical symmetric cosmological solutions in TMG.\\
Looking at \eqref{eq:Horizon} one can see that the entropy \eqref{eq:TMGThermalEntropy}, similar to the Einstein gravity case, corresponds to non-spherically symmetric cosmological solutions in TMG since $\beta$ is still allowed to fluctuate. So taking into account the result obtained for the entropy \eqref{eq:TMGThermalEntropy} one can conclude that also in TMG our boundary conditions describe a certain class of non-spherically symmetric solutions in flat space whose area is the same as for spherically symmetric solutions.

%---------------------------------------------------------------------------------------------------------------
\section{Relation with Troessaert BCs}\label{eq:Troessaert}
%---------------------------------------------------------------------------------------------------------------

In this section we show that the boundary conditions presented in this paper can also be related to the boundary conditions proposed in \cite{Troessaert:2013fma} via sending the cosmological constant to zero. On the level of the metric this can be seen by first rewriting the boundary conditions (plus additional subleading contributions) of \cite{Troessaert:2013fma} in BMS gauge\footnote{Please note the absence of the terms linear in $x^\pm$ in the conformal factor that are present in \cite{Troessaert:2013fma}.} as
    \begin{subequations}
    \begin{align}
		g_{rr}&=0,\\
		g_{rx^+}&=-\frac{\ell}{2}e^{\phi},\\
		g_{rx^-}&=-\frac{\ell}{2}e^{\bar{\phi}},\\
		g_{x^+x^+}&=-\ell^2\mathcal{L},\\
		g_{x^+ x^-}&=-\frac{e^{\phi+\bar{\phi}}}{2}\left[r^2+\ell^2\left(e^{-2\phi}\mathcal{L}+e^{-2\bar{\phi}}\bar{\mathcal{L}}\right)\right],\\
		g_{x^-x^-}&=-\ell^2\bar{\mathcal{L}},
    \end{align}
    \end{subequations}
where all unbarred quantities are functions of $x^+$, all barred quantities are functions of $x^-$, $x^\pm=\frac{t}{\ell}\pm\varphi$ and $\ell$ is the AdS radius.\\
Introducing new functions $\mathcal{M}$, $\mathcal{N}$, $\alpha$ and $\beta$ as
    \begin{subequations}
    \begin{align}
        \mathcal{L}=&-\frac{1}{4}\left(\mathcal{M}+2\epsilon\mathcal{N}\right),&\phi=&\alpha+\epsilon\beta,\\
        \bar{\mathcal{L}}=&-\frac{1}{4}\left(\mathcal{M}-2\epsilon\mathcal{N}\right),&\phi=&\alpha-\epsilon\beta,
    \end{align}
    \end{subequations}
as well as replacing $t\rightarrow u$ one obtains the following metric:
    \begin{subequations}
    \begin{align}
		g_{rr}&=0,\\
		g_{r\varphi}&=-\ell e^{\alpha}\sinh[\tfrac{\beta}{\ell}],\\
		g_{ru}&=-e^{\alpha}\cosh[\tfrac{\beta}{\ell}],\\
		g_{\varphi\varphi}&=e^{2\alpha}r^2-\ell^2\left(\mathcal{M}\sinh^2[\tfrac{\beta}{\ell}]-\frac{\mathcal{N}}{\ell}\sinh[\tfrac{2\beta}{\ell}]\right),\\
		g_{\varphi u}&=\mathcal{N},\\
		g_{uu}&=e^{2\alpha}\frac{r^2}{\ell^2}+\mathcal{M}\cosh^2[\tfrac{\beta}{\ell}]-\frac{\mathcal{N}}{\ell}\sinh[\tfrac{2\beta}{\ell}].
    \end{align}
    \end{subequations}
In the limit $\ell\rightarrow\infty$ one then exactly recovers the boundary conditions \eqref{eq:BCsMetric}. Of course a similar procedure as outlined in e.g. \cite{Krishnan:2013wta,Riegler:2016hah} can be used in the Chern-Simons Formalism reproducing exactly \eqref{eq:NewFSBCs}.\\
One can also perform such a limiting procedure at the level of asymptotic symmetry algebras. Taking two copies of a semi-direct product of a Virasoro algebra (with generators $\mathfrak{L}_n$ and $\bar{\mathfrak{L}}_n$) and an affine $\hat{\mathfrak{u}}(1)$ current algebra (with generators $\mathfrak{J}_n$ and $\bar{\mathfrak{P}}_n$) with central charges $c=\bar{c}=\frac{3\ell}{G_N}$ and $\hat{\mathfrak{u}}(1)$ levels $\kappa=\bar{\kappa}=-\frac{\ell}{8G_N}$ it is straightforward to show that an after the redefinitions
    \begin{subequations}
    \begin{align}
        L_n:=&\mathfrak{L}_n-\bar{\mathfrak{L}}_{-n},&M_n:=&\frac{1}{\ell}\left(\mathfrak{L}_n+\bar{\mathfrak{L}}_{-n}\right),\\
        J_n:=&\mathfrak{J}_n-\bar{\mathfrak{J}}_{-n},&P_n:=&\frac{1}{\ell}\left(\mathfrak{P}_n+\bar{\mathfrak{P}}_{-n}\right),
    \end{align}
    \end{subequations}
one precisely obtains \eqref{eq:ASAMiddle} in the limit $\ell\rightarrow\infty$.

%---------------------------------------------------------------------------------------------------------------
\section{Conclusion}
%---------------------------------------------------------------------------------------------------------------

In this paper we presented a new set of boundary conditions for asymptotically flat spacetimes in 2+1 dimensions. We determined the asymptotic symmetries for these boundary conditions for both Einstein gravity as well as TMG. For Einstein gravity we found a semi-direct product of the $\mathfrak{bms}_3$ algebra with two $\hat{\mathfrak{u}}(1)$ current algebras. The corresponding central charge $c_M$ and $\hat{\mathfrak{u}}(1)$ level $\kappa_P$ were found to be $c_M=\frac{1}{4G_N}$ and $\kappa_P=-\frac{1}{4G_N}$ in the Einstein gravity case. For TMG we recovered the expected deformation $c_L=\frac{3}{\mu G_N}$ and $\kappa_J=-\frac{3}{8\mu G_N}$.\\
Then looking at the thermal entropy via calculating the horizon area of these new solutions we found that even though the horizon does not necessarily have to be spherically symmetric the thermal entropy still coincides with the expression one would expect from an ordinary spherically symmetric cosmological solution in flat space. We then confirmed this result by determining the thermal entropy also using the first law of flat space cosmologies as well as appropriate holonomy conditions and showed that this approach is equivalent to a Wilson line wrapping around the horizon.\\
As a final point we also showed how our boundary conditions can be interpreted from a point of view of a limit of vanishing cosmological constant for the Troessaert boundary conditions \cite{Troessaert:2013fma}.\\
Even though our boundary conditions allow for non-spherically symmetric cosmological solutions they only allow for horizon deformations that do not change the total area of a spherically symmetric FSC. It would thus be interesting to investigate if there are further generalizations of our boundary conditions that allow for horizon deformations that change the area of the horizon. One possible way to look for extensions like this could be (non-principally embedded) flat space higher-spin gravity \cite{Afshar:2013vka, Gonzalez:2013oaa}. Furthermore it could be of interest to use the approach outlined in \cite{Bagchi:2014iea,Basu:2015evh,Riegler:2016hah} to determine holographic entanglement entropy using our boundary conditions.\\
Another possible interesting extension would be to apply our boundary conditions to other higher-derivative theories besides TMG like e.g. New Massive Gravity \cite{Bergshoeff:2009hq} as well as extensions to dimensions higher that three.\\

%---------------------------------------------------------------------------------------------------------------
\subsection*{Acknowledgments}
%---------------------------------------------------------------------------------------------------------------

We are grateful to Martin Ammon, Glenn Barnich, Geoffrey Comp{\`e}re and Daniel Grumiller for enlightening discussions and comments. In addition MR wants to thank Martin Ammon and the TPI at the Friedrich-Schiller-Universit{\"a}t Jena for the opportunity of an extended visit during the final stages of this project. SD is a Research Associate of the Fonds de la Recherche Scientifique F.R.S.-FNRS (Belgium). He is supported in part by the ARC grant “Holography, Gauge Theories and Quantum Gravity Building models of quantum black holes”, by IISN - Belgium (convention 4.4503.15) and benefited from the support of the Solvay Family. The research of MR is supported by the ERC Starting Grant 335146 "HoloBHC".

%---------------------------------------------------------------------------------------------------------------
% BIBLIOGRAPHY
%---------------------------------------------------------------------------------------------------------------

\providecommand{\href}[2]{#2}\begingroup\raggedright\endgroup

\end{document}